\documentclass[preprint,aps,showpacs,superscriptaddress]{revtex4}

\usepackage{epsfig,pstricks}
\usepackage{amsmath,amssymb}

\def\CC{{\rm\kern.24em \vrule width.04em height1.46ex depth-.07ex
\kern-.30em C}}
\def\RR{{\rm
         \vrule width.04em height1.58ex depth-.0ex
         \kern-.04em R}}
\def\id{{\rm 1\kern-.22em l}}

\def\tr{\mathrm{tr}}
\newcommand{\beq}{\begin{equation}}
\newcommand{\beqa}{\begin{eqnarray}}
\newcommand{\nbeqa}{\begin{eqnarray*}}
\newcommand{\eeq}{\end{equation}}
\newcommand{\eeqa}{\end{eqnarray}}
\newcommand{\neeqa}{\end{eqnarray*}}
\newcommand{\bra}[1]{\left\langle #1 \right |}
\newcommand{\ket}[1]{\left | #1 \right\rangle}
\newcommand{\braket}[2]{\left\langle #1 | #2 \right\rangle}

\begin{document}

\title{Entangled three-qubit states without concurrence and three-tangle}

\author{Robert Lohmayer} 
\affiliation{Institut f\"ur Theoretische Physik, 
         Universit\"at Regensburg, D-93040 Regensburg, Germany}
\author{Andreas Osterloh}
\affiliation{Institut f\"ur Theoretische Physik, 
         Universit\"at Hannover, D-30167 Hannover, Germany}
\author{Jens Siewert}
\affiliation{MATIS-INFM $\&$ Dipartimento di Metodologie Fisiche e
    Chimiche (DMFCI), viale A. Doria 6, 95125 Catania, Italy}
\affiliation{Institut f\"ur Theoretische Physik, 
         Universit\"at Regensburg, D-93040 Regensburg, Germany}
\author{Armin Uhlmann}
\affiliation{Institut f\"ur Theoretische Physik, 
         Universit\"at Leipzig, D-04109 Leipzig, Germany
            }


\begin{abstract}
We provide a complete analysis of mixed three-qubit states 
composed of a GHZ state and a W state orthogonal to the former.
We present optimal decompositions and convex roofs
for the three-tangle. Further, we provide an analytical method 
to decide whether or not an arbitrary rank-2 state of three qubits
has vanishing three-tangle.
These results highlight intriguing differences compared to the
properties of two-qubit mixed states, and 
may serve as a quantitative reference for future
studies of entanglement in multipartite mixed states.
By studying the 
Coffman-Kundu-Wootters inequality  we find that,
while the amounts of inequivalent entanglement types strictly add
up for pure states, 
this ``monogamy'' can be lifted for mixed states
by virtue of vanishing tangle measures.
\end{abstract}
\maketitle

Characterizing and quantifying entanglement of multipartite mixed states is
a fundamental issue in quantum information theory, both from
a theoretical and a practical point of view~\cite{Werner1989,Bennett1996,BennettDiVincenzo1996}.  
Despite the large number of profound 
results, 
e.g.~\cite{Duer1999,Coffman2000,Duer2000,Acin2000,Carteret2000,Acin2001,AcinBruss2001,Wei2003,Levay2005,Yu2006}, 
even for the simplest case -- the one of three qubits --
there is no general solution to this problem. 
While the entanglement problem is well understood for pure and mixed states
of two qubits~\cite{Werner1989,Bennett1996,BennettDiVincenzo1996,Wootters1998,Uhlmann2000,Wootters2001}, 
expanding our knowledge to both higher-dimensional systems and systems
of more than two qubits turned out to be a formidable task.
Although pure three-qubit states
are well characterized by their local invariants and their
non-trivial entanglement properties, similarly complete understanding
of mixed three-qubit states remains elusive.

In Ref.~\cite{Uhlmann2000} it has been worked out that mixed states
are characterized in terms of the convex roof for pure-state entanglement
measures. The seminal paper by Coffman et al.~\cite{Coffman2000} provided
a basis for the quantification of three-party entanglement
by introducing the so-called residual tangle that led to 
understanding important
subtleties in multipartite quantum correlations~\cite{Duer2000}.
To date, there is no simple analytical method to compute the residual
tangle for mixed three-qubit states. Therefore, it may be helpful to
study simple cases of such mixed states which, on the one hand, allow
for a complete characterization, but on the other hand involve as many
features as possible that go beyond the well-known bipartite case.

To this end, we study a particular family of rank-2 mixed three-qubit states.
After briefly introducing the basic concepts to describe two-qubit and
three-qubit entanglement we first analyze superpositions of  GHZ and
 W states. Then we present the results for
the residual tangle of GHZ/W mixtures that lead us to 
the optimal decompositions, and the corresponding convex roofs. 
Surprisingly, we obtain the  residual tangle not only
for mixtures of GHZ and W states, but also for a large part of the
Bloch sphere spanned by these states.
Finally we discuss the Coffman-Kundu-Wootters (CKW) inequality. 
Our results provide a rich reference for 
further work on the quantification of multipartite 
entanglement.

{\em Important concepts. --}
The concurrence $C(\phi_{AB})$ measures the degree of bipartite entanglement
shared between the parties $A$ and $B$ in a pure two-qubit state
$\ket{\phi_{AB}}\in{\cal H}_A\otimes{\cal H}_B$. In terms of the coefficients
$\{\phi_{00}, \phi_{01}, \phi_{10}, \phi_{11}\}$ of $\ket{\phi_{AB}}$
with respect
to an orthonormal basis it is defined as
\begin{equation}
 C\ =\ 2 | \phi_{00}\phi_{11}-  \phi_{01}\phi_{10} |
  \ \ .
\label{def_concurrence}
\end{equation}
The concurrence is maximal for Bell states like
\mbox{$\frac{1}{\sqrt{2}}(\ket{00}+\ket{11})$} and vanishes 
for factorized states.

The 3-tangle (or residual tangle) $\tau_3(\psi)$,
a measure for three-party entanglement  in the three-qubit state
$\ket{\psi}\in {\cal H}_A\otimes{\cal H}_B\otimes{\cal H}_C$
has been introduced in
Ref.~\cite{Coffman2000}. 
It can be expressed by using the wavefunction coefficients
$\{\psi_{000},\psi_{001},\ldots,\psi_{111}\}$ as
\beqa
\label{def_3tangle}
\tau_3 &=& 4\ |d_1 - 2d_2 + 4d_3|\\
  d_1&=& \psi^2_{000}\psi^2_{111} + \psi^2_{001}\psi^2_{110} + \psi^2_{010}\psi^2_{101}+ \psi^2_{100}\psi^2_{011} \nonumber \\
  d_2&=& \psi_{000}\psi_{111}\psi_{011}\psi_{100} + \psi_{000}\psi_{111}\psi_{101}\psi_{010}\nonumber \\ 
    &&+ \psi_{000}\psi_{111}\psi_{110}\psi_{001} + \psi_{011}\psi_{100}\psi_{101}\psi_{010}\nonumber \\
    &&+ \psi_{011}\psi_{100}\psi_{110}\psi_{001} + \psi_{101}\psi_{010}\psi_{110}\psi_{001}\nonumber \\
  d_3&=& \psi_{000}\psi_{110}\psi_{101}\psi_{011} + \psi_{111}\psi_{001}\psi_{010}\psi_{100}\nonumber \ \ .
\eeqa
We mention that this definition of the 3-tangle coincides with the modulus
of a so-called hyperdeterminant which was introduced by 
Caley~\cite{Caley1845,Miyake2003}.
For the GHZ state
\begin{equation}
       \ket{GHZ}\ = \ \frac{1}{\sqrt{2}}\ (\ket{000}\ +\ \ket{111})
\label{ghz}
\end{equation}
the 3-tangle becomes maximal: $\tau_3(GHZ)=1$, and it
vanishes for any factorized state. Remarkably there is a class of
entangled three-qubit states for which $\tau_3$ vanishes~\cite{Duer2000}.
This class is represented by the $\ket{W}$ state
\begin{equation}
  \ket{W}\ =\ \frac{1}{\sqrt{3}}\ ( \ket{100}\ +\ \ket{010}\ +\ \ket{001} )
  \ \ .
\label{wstate}
\end{equation}
Now consider, e.g., mixed two-qubit states $\omega$
with their decompositions
\begin{equation}
   \omega\ =\ \sum_j p_j \pi_j,\ \ 
                                \ \ \ \pi_j=\frac{\ket{\phi_j}\!\bra{\phi_j}}
                                        {\langle \phi_j | \phi_j \rangle}\ \ .
\end{equation}
The concurrence of the mixed two-qubit state $\omega$
is defined as the average pure-state concurrence minimized over
all decompositions
\begin{equation}
  C(\omega)\ =\ \min \sum p_j C(\pi_j)\ \ .
\label{mixed_concurrence}
\end{equation}
which is also called a convex-roof extension~\cite{BennettDiVincenzo1996,Benatti1996,Uhlmann2000}.
The mixed-state 3-tangle is defined analogously.
A decomposition that realizes the minimum of the respective function,
is called optimal.

In this paper we study the 3-tangle of rank-2 mixed three-qubit states
\beq
    \rho(p)\ =\ p\ \pi_{\mathrm{GHZ}} \ +\ (1-p)\ \pi_{\mathrm W}
\label{mixedstate}
\eeq
with $\pi_{\mathrm{GHZ}}=\ket{GHZ}\!\bra{GHZ}$ and
    $\pi_{\mathrm W}=\ket{W}\!\bra{W}$ (notice that $\braket{GHZ}{W}=0$).

{\em Superpositions of GHZ and W states. --}
It is well-known that from the decomposition of a rank-$n$ density matrix
$\rho=\sum_{j=1}^{n} p_j \ket{j}\!\bra{j}$
into its eigenstates $\{\ket{j}\}$,
any other decomposition $\rho=\sum \ket{\chi_l}\!\bra{\chi_l}$
of length $m \ge n$ can be obtained with a unitary 
\mbox{$m\times m$} matrix $U_{lj}$ 
via $\ket{\chi_l}\ = \ \sum_{j=1}^m U_{lj} (\sqrt{p_j}\ket{j})$.
Hence, the vectors of any decomposition of our states
$\rho(p)$ are linear combinations of $\ket{GHZ}$ and $\ket{W}$.
Therefore we first analyze the 3-tangle of the states
\beq
   \ket{Z(p,\varphi)}\ =\ \sqrt{p}\ket{GHZ}\ -\ e^{i\varphi}\sqrt{1-p}\ket{W}
        \ \ .
\label{zstate}
\eeq
With Eq.~(\ref{def_3tangle}) we obtain (see also Fig.~1)
\beq
     \tau_3(Z(p,\varphi))\ =
     \ \left|p^2-\frac{8\sqrt{6}}{9}\sqrt{p(1-p)^3}e^{3i\varphi}\right|\ \ .
\label{3tangle_zstate}
\eeq
\begin{figure}
\resizebox{.4\textwidth}{!}{\includegraphics{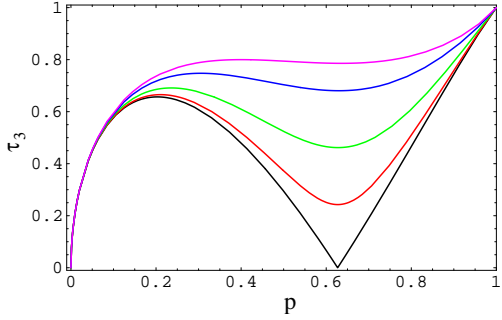}}
\caption{The 3-tangle of the states $\ket{Z(p,\varphi)}$ in
         Eq.~(\ref{zstate}) for various values of $\varphi=\gamma \cdot
         \frac{2\pi}{3}$ (from top to bottom: $\gamma=1/2,1/3,1/5,1/10,0$).
The 3-tangle 
is periodic in $\varphi$ with a period of $2\pi/3$.
Further, $\tau_3(Z)$ vanishes trivially for zero GHZ component, 
but also for $\varphi=0$ and non-zero GHZ amplitude
      $p_0\ =\ \frac{4\sqrt[3]{2}}{3+4\sqrt[3]{2}} \ \approx\ 0.627$.
This is in striking contrast to the two-qubit case where,
in a superposition of a Bell state with an orthogonal unentangled state,
the concurrence equals the weight of the Bell state~\cite{Abouraddy2001},
that is the entanglement remains ``untouched''.
      }
\end{figure}
%
We notice that $\tau_3(Z)$ has a non-trivial zero ($\varphi=0$) 
\[
            p_0\ =\ \frac{4\sqrt[3]{2}}{3+4\sqrt[3]{2}} \ =\ 0.626851\ldots
\]

{\em Mixtures of GHZ and W states. --}
An immediate consequence of the existence of
a finite $p_0$ with $\tau_3(Z(p_0))=0$, together with the permutation symmetry
of the states, is
that the mixed-state 
3-tangle $\tau_3(\rho(p))=0$ for $p=p_0$ as well as for all $0\le p\le p_0$.
For $p=p_0$ we have the optimal decomposition 
\beqa
   \rho(p_0)\ & = &\ \frac{1}{3}(
             \ket{Z_0^{0}}\!\bra{Z_0^{0}}+
             \ket{Z_0^{1}}\!\bra{Z_0^{1}}+
             \ket{Z_0^{2}}\!\bra{Z_0^{2}})
   \label{decomp_p0}
   \\
   \ket{Z_0^{j}}\ & = &\ \sqrt{p_0}\ket{GHZ}\ -\ e^{\frac{2\pi i}{3}j}
                   \sqrt{1-p_0}\ket{W}\ \ .
   \nonumber
\eeqa
For $p<p_0$ there is a decomposition with vanishing 3-tangle of the form
$ \rho(p) = \frac{p}{p_0} \rho(p_0) +
                        \left(1-\frac{p}{p_0}\right) \pi_{\mathrm{W}}$.
For $p>p_0$, all states
$\rho(p)$ have non-vanishing 3-tangle.

The decomposition in Eqs.~(\ref{decomp_p0})
provides a 
trial decomposition $\cal S$ for $\rho(p>p_0)$ given by 
${\cal S}=\{\sqrt{\frac{1}{3}}\ket{Z^0},\sqrt{\frac{1}{3}}\ket{Z^1},\sqrt{\frac{1}{3}}\ket{Z^2}\}$ with
$\ket{Z^j}=\ket{Z(p,\frac{2\pi}{3}j)}$.
Its 3-tangle 
\beq
       g_{\mathrm{I}}(p)=
                    p^2-\frac{8\sqrt{6}}{9}\sqrt{p(1-p)^3}\ ,\ \ p\ge p_0
\eeq
provides an upper bound for $\tau_3(\rho(p>p_0))$:
as the function $g_{\mathrm{I}}(p)$ is not convex 
in the entire interval $p_0\le p \le 1$, it cannot represent 
the 3-tangle of $\rho(p)$ for all $p\in [p_0,1]$.

In order to test this upper bound, we have performed extensive
numerical studies (the numerical method can be
tested with the two-qubit case as well as with states $\rho(p\le p_0)$ 
with vanishing 3-tangle). According to Caratheodory's theorem, 
for rank-2 states
four vectors are sufficient to minimize the 3-tangle. Hence
we need to investigate decompositions with 2, 3, or 4 vectors only.

Our numerical results indicate that $\cal S$ is indeed an optimal 
decomposition for values of $p$ close to $p_0$. For larger $p$, however,
we propose the four-vector decomposition $\cal T$ that consists
of three vectors analogous to $\cal S$ and a {\em pure}
GHZ state:
\beqa
   \rho(p) & =  (1-b)\  &\ket{GHZ}\!\bra{GHZ}   + 
             \  \frac{b}{3}\ \sum_{j=0}^2 \ket{Z^{j}(a)}\!\bra{Z^{j}(a)}
   \nonumber
\eeqa
where $\ket{Z^{j}(a)}\equiv \ket{Z(a,\frac{2\pi}{3}j)}$.
The coefficients $a$ and $b$ have to be determined as functions of $p$.
It turns out that the decomposition $\cal S$ is optimal for 
$p_0\le p \le p_1 = \frac{1}{2} + \frac{3}{310}\sqrt{465}\equiv 0.70868\ldots$
For $p\in [p_1,1]$ the decomposition $\cal T$ has the lowest
average 3-tangle, given by the linear function
\beq
    g_{\mathrm{II}}(p)\ =\ 1\ -\ (1-p)\left(\frac{3}{2}+\frac{1}{18}
                                       \sqrt{465}
                                   \right)\ \ .
\label{tangle_decomp_T}
\eeq
This is in complete agreement with the numerical results.

{\em Convex roofs. --}
With these findings
we can elucidate the affine regions (the convex roofs) of the
3-tangle for any mixture of GHZ and W states.
To this end, imagine the two-dimensional space spanned by the GHZ and the
W state represented by a Bloch sphere with the GHZ state at
its north pole and the W state at the south pole (cf.~Fig.\ 2).
\begin{figure}
\resizebox{.4\textwidth}{!}{\includegraphics{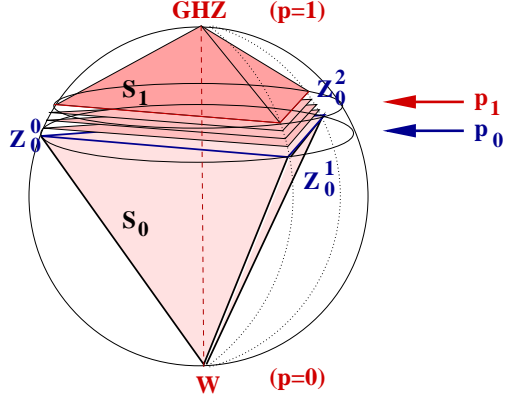}}
\caption{Bloch sphere for the two-dimensional space spanned
         by the GHZ state and the W state.
         The simplex $S_0$ contains all states with zero 3-tangle.
         The ``leaves'' between $p_0$ and $p_1$ represent sets of
         constant 3-tangle, and in the simplex $S_1$ the 3-tangle
         is affine (for further explanation, see text).
      }
\end{figure}
All the states $\ket{Z(p,\varphi)}$ are located at the unit sphere, with
azimuth $\varphi$ and $p$ the distance from the south pole
along the north-south line (i.e., $p=0$ is the south pole,
and $p=1$ is the north pole).
The states with zero 3-tangle are represented by the simplex $S_0$
with corners $W$, $Z_0^0$, $Z_0^1$, $Z_0^2$. 

Now consider planes $P(p)$
parallel to the ground plane of the simplex (i.e., the plane
containing the triangle $Z_0^0,Z_0^1,Z_0^2$) and intersection
point $p$ with the north-south line.
For $p_0\le p\le p_1$ the 3-tangle is constant in a triangle
that has its corners at the intersection points of the plane $P(p)$
and the meridians through $Z_0^0$, $Z_0^1$, and $Z_0^2$ (see Fig.~2).

For $p_1\le p\le 1$ we have another simplex $S_1$ with affine 3-tangle.
The ground plane of this simplex is formed by the last of the ``leaves''
(at $p=p_1$) described above, and the top corner is the GHZ state. 
That is,
in each plane parallel to the ground plane of this
simplex the 3-tangle is constant. The 3-tangle of any point
inside $S_1$ (with distance $p^{\prime}$ from the south pole, i.e.,
the GHZ weight equals $p^{\prime}$) is a convex combination 
$\alpha \cdot g_{\mathrm{II}}(p_1) + \beta\cdot 1$
of $g_{\mathrm{II}}(p_1)$
(the 3-tangle in the ground plane of $S_1$)
and of 1, the value for the GHZ state. The coefficients are
$\alpha=\frac{1-p^{\prime}}{1-p_1}$ and
$\beta=\frac{p^{\prime}-p_1}{1-p_1}$. 
This completes the characterization of the mixed states $\rho(p)$.

{\em CKW inequality. --}
Given a whole family of mixed three-qubit states with the corresponding
3-tangle one might like to check the CKW relations~\cite{Coffman2000}.
To this end, we consider 
pure three-qubit states
$\ket{\psi}\in {\cal H}_A\otimes{\cal H}_B\otimes{\cal H}_C$
and introduce the reduced two-qubit density matrices
$\rho_{AB}=\tr_C(\ket{\psi}\!\bra{\psi})$, 
$\rho_{AC}=\tr_B(\ket{\psi}\!\bra{\psi})$, and 
the reduced density matrix of the first qubit
$\rho_A=\tr_{BC}(\ket{\psi}\!\bra{\psi})$. For pure states,
one has the ``monogamy relation''
$
    4 \det(\rho_A)\ = \ C(\rho_{AB})^2+C(\rho_{AC})^2+\tau_3(\psi)
$, that is, the entanglement
of  qubit $A$ can be either bipartite entanglement with $B$ or $C$
(concurrence),
or three-party entanglement with $B$ {\em and} $C$ (3-tangle).
For mixed three-qubit states $\rho$, Coffman {\em et al.} found the inequality
\beq
  4  \min \left(\det(\rho_A)\right)\ \ge\ C(\rho_{AB})^2+C(\rho_{AC})^2\ \ .
\label{ckw_inequality}
\eeq
Here, also the 1-tangle has to be minimized for all possible
decompositions of $\rho$. 
The reduced two-qubit density matrices are defined analogously as
$\rho_{AB}=\tr_C(\rho)$, 
$\rho_{AC}=\tr_B(\rho)$, 
$\rho_{A}=\tr_{BC}(\rho)$. 
We refer to Eq.~(\ref{ckw_inequality})
as the ``CKW inequality''.

Due to the invariance of $\rho(p)$ with respect to qubit permutations,
 we can consider the CKW inequality 
with respect to the first qubit, without loss of generality.
It is straightforward to compute the concurrences:
\begin{equation}
    C_{AB}^2 + C_{AC}^2 =  2 \left(\max{\left[0,
                                 \frac{2}{3}(1-p)-\sqrt{\frac{p}{3}(2+p)}
                                         \right]}
                              \right)^2\ \ .
\end{equation}
This is a monotonously decreasing function for $p\in [0,1]$
that vanishes for $p_C=7-\sqrt{45}\approx 0.292$.
For the minimum 1-tangle we obtain
\beq
    4 \min\ \det \left(\mathrm{tr}_{BC} \rho(p)\right)\ = \
             \frac{1}{9}\ (\ 5p^2\ -\ 4p\ +\ 8\ )\ \ .
\eeq

In order to interpret these results,  
we plot the 1-tangle, the sum of squared concurrences, 
and the 3-tangle, see Fig.~3. 
\begin{figure}[b]
\resizebox{.4\textwidth}{!}{\includegraphics{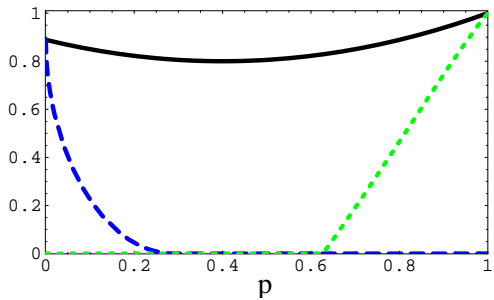}}
\caption{   Plot of the 1-tangle (solid line), the sum of squared
            concurrences $C(\rho_{AB})^2+C(\rho_{AC})^2$ (dashed line)
            and 3-tangle (dotted line) in $\rho(p)$ as a function of $p$.
            Clearly, the CKW inequality~(\ref{ckw_inequality}) is satisfied.
      }
\end{figure}
 The 1-tangle is always larger than the sum of squared concurrences and
 3-tangle. In particular, 
there is a region $p_C \le p \le p_0$
where there is quite substantial  1-tangle, but there is {\em zero}
concurrence and {\em zero} 3-tangle in the state $\rho(p)$. Our interpretation
of this result
is that it is not pre-determined whether the entanglement contained
in $\rho(p)$ is bipartite entanglement or three-way entanglement.
According
to the choice of decomposition
the entanglement in $\rho(p)$ can be represented either
in terms of bipartite correlations, or as three-partite correlations,
respectively. This is a curious fact if we recall that the 
GHZ and the W state are
locally inequivalent.  Apparently, for mixed states
there is no strict monogamy as for pure states.
As opposed to this, for ``less mixed states'' -- i.e., for
$p<p_C$ and for $p > p_0$ --
the entanglement stems (mandatorily) to a large extend
either from bipartite correlations, or from three-way correlations.

{\em Conclusion. --}
Summarizing, we have studied the entanglement properties of the family
of mixed three-qubit states $\rho(p)$ (cf.~Eq.~(\ref{mixedstate})).
As opposed to the two-qubit case (where one has a huge variety
of optimal decompositions~\cite{Wootters1998,Uhlmann2000}), 
for our three-qubit states $\rho(p)$ there appears to be only a single type
of optimal decomposition (apart from phase rotations). Further,
while a two-qubit state of rank $n$ has always an optimal decomposition
of length $n$, this does not hold for the three-qubit states
$\rho(p)$. 
Remarkably, we have obtained the convex roof of the 3-tangle not only
for mixtures as in~Eq.~(\ref{mixedstate}), but for a considerable part
of the Bloch sphere (cf.~Fig.~2).
We mention that, if any of these density matrices is convexly combined
with an {\em arbitrary} three-qubit density matrix, our results provide a 
non-trivial upper bound for the 3-tangle of the resulting state.

Moreover, we have found that for the entanglement contained
in multipartite mixed states there is no strict monogamy, i.e., it
can be represented by different types of locally inequivalent 
quantum correlations.

Finally, we would like to point out that from our results we may draw an important
conclusion for arbitrary rank-2 density matrices $\rho=p \ket{1}\!\bra{1}+(1-p)\ket{2}\!\bra{2}$, 
of three qubits. Also in the  general case, there
will exist a simplex $S_0$ which makes it possible to decide on analytical
grounds whether or not a rank-2 state has vanishing 3-tangle. This can be seen as
follows. 

Consider the 3-tangle of the state \mbox{$\ket{\psi}=\ket{1}+z \ket{2}$} with
a complex number $z$ (normalization is irrelevant here). The condition for the 3-tangle
of this state to be zero (analogously to Eq.~(\ref{3tangle_zstate})) is given by a polynomial
equation of 4th degree in $z$. Hence, there are four (in general different) pure states with vanishing
3-tangle that define the simplex $S_0$ in the corresponding Bloch sphere 
which contains those density matrices whose 3-tangle is zero. Thus, the 3-tangle of $\rho$
vanishes {\em iff} it belongs to $S_0$.

The implications of the concept of the simplex $S_0$ are reaching even further. It can be
generalized for multipartite entanglement monotones of
 pure states~\cite{Vidal2000,VerstraeteDM2003,BriandLT2003}. 
A method to explicitly write $N$-qubit entanglement monotones ($N\ge 3$) as certain expectation values 
of antilinear operators has been presented in Refs.~\cite{Osterloh2005,Osterloh2006}.
Let us consider an arbitrary rank-2 $N$-qubit density matrix.
In complete analogy to what was said above, one can find the zeros of such $N$-tangles on
the Bloch sphere which then define a ``zero-polyhedron'' (rather than a simplex) that contains
all the density matrices with vanishing $N$-tangle.

{\em Acknowledgment. --}
The authors would like to thank L.\ Amico and A.\ Fubini
for stimulating discussions. This work was supported by the EU
RTN grant HPRN-CT-2000-00144
and the Sonderforschungsbereich 631
of the German Research Foundation. JS. receives support from
the Heisenberg Programme of the German Research Foundation.
%


\end{document}